\documentclass{article}
\usepackage{spconf,amsmath,graphicx,hyperref}

\title{Towards Personalized, Precise and Survey-free Environment Recognition: AI-Enhanced Sensor Fusion without Pre-Deployment}
\name{Ruichen Wang, Zhikang Ni, Pengzhou Wang, Xiya Cao, Zhi Li, Bao Zhang}
\address{Huawei Technologies Co. Ltd}
\begin{document}
%\ninept
%
\maketitle
\begin{abstract}
Accurate and personalized environment recognition is essential for seamless indoor positioning and optimized connectivity, yet traditional fingerprinting requires costly site surveys and lacks user-level adaptation. We present a survey-free, on-device sensor-fusion framework that builds a personalized, lightweight multi-source fingerprint (FP) database from pedestrian dead reckoning (PDR), WiFi/cellular, GNSS, and interaction time tags. Matching is performed by an AI-enhanced dynamic time warping module (AIDTW) that aligns noisy, asynchronous sequences. To turn perception into continually improving actions, a cloud-edge online Reinforcement Learning from Human Feedback (RLHF) loop aggregates desensitized summaries and human feedback in the cloud to optimize a policy via proximal policy optimization (PPO), and periodically distills updates to devices. Across indoor/outdoor scenarios, our system reduces network-transition latency (measured by time-to-switch, TTS) by 32-65\% in daily environments compared with conventional baselines, without site-specific pre-deployment.
\end{abstract}
\section{Introduction}
\label{sec:intro}

Timely and accurate recognition of user environments (e.g., corridor, stairwell, restroom, outdoors) underpins seamless indoor positioning and adaptive connectivity. Conventional fingerprinting—built on site surveys and static radio maps—does not scale and adapts poorly to user routines, device heterogeneity, and temporal drift, leading to slow responses during short transitions that dominate perceived latency.

We propose a personalized, survey-free framework that constructs a multi-source fingerprint library from signals produced in daily use. The system fuses PDR, WiFi/cellular indicators, opportunistic GNSS, and interaction time tags into compact temporal fingerprints, enabling a personalized, lightweight on-device database. Matching is performed by an AI-enhanced dynamic time warping module, AIDTW, that aligns noisy, asynchronous sequences. To translate perception into actions that continually improve, a cloud-edge online RLHF loop aggregates desensitized summaries and human feedback in the cloud to optimize a policy and periodically distills updates to devices. This collaboration lets the system update itself over time without site-specific surveys.

\textbf{Contributions.} This paper makes three contributions:
\begin{itemize}
  \item \textbf{Personalized multi-source fingerprint database:} passive, survey-free fusion that yields a personalized, lightweight database on device.
  \item \textbf{AI-enhanced matching via AIDTW:} robust cross-modal alignment that enables 32-65\% faster network switching in daily environments.
  \item \textbf{Cloud-edge online RLHF for policy optimization:} privacy-preserving training and distillation that keeps policies improving over time.
\end{itemize}

\section{Methodology}
\subsection{Related work}

Classical WiFi/Cellular fingerprinting builds site-specific radio maps and matches online RSS/CSI via nearest neighbors or probabilistic models \cite{C1,C2,C3}. Deep models and CSI features improve robustness but still inherit survey and maintenance costs \cite{C16,C4,C14}. To lower surveying effort, self-updating maps, crowdsourcing, and interpolation have been explored \cite{C5,C13,C15}. Beyond site surveys and static radio maps, prior work explored unsupervised and crowd-sourced mapping to remove war-driving, but drift and venue-specific priors remained challenging \cite{C18}.

Hybrid systems combine inertial PDR \cite{RIO2022} with radio and geomagnetic cues to stabilize room- or zone-level recognition \cite{C13,C12}. Prior techniques typically require anchor deployment, manual calibration, or venue-specific priors. We instead normalize heterogeneous signals (PDR, WiFi/Cellular, GNSS, time tags) into compact temporal fingerprints that can be collected passively and updated per user.

Dynamic Time Warping (DTW) remains a strong baseline for aligning rate-mismatched sequences \cite{C7}. Differentiable relaxations (Soft-DTW) enable metric learning and end-to-end optimization \cite{C6}. Our AIDTW couples neural similarity with DTW, applies adaptive temporal weights, and uses dropout-robust path constraints to align noisy, partially observed multi-modal traces in real time\cite{C19}.

Policy optimization methods such as PPO provide stable on-policy training for resource-constrained control \cite{C11}. RLHF further aligns learned policies with human preferences and corrections \cite{C10}. We integrate PPO with online RLHF in a cloud-edge collaboration: the cloud aggregates desensitized summaries and feedback, learns a reward model and policy, and periodically distills updates to devices\cite{C8,C9}, enabling the system to update itself over time \cite{C20}.

Compared with prior art, our system (i) eliminates site surveys by building a personalized, lightweight multi-source fingerprint database from passive signals, (ii) introduces AIDTW for robust cross-modal alignment under asynchrony and missing data, and (iii) closes the perception–action loop with a cloud-edge online RLHF framework (PPO + periodic distillation) that updates policies over time from human feedback.

\subsection{Personalized multi-source data collection and fingerprint generation}
\label{sec:collection}

We build a personalized and lightweight fingerprint from signals passively collected on commodity devices: WiFi (BSSID, RSSI), cellular (Cell ID, RSRP/RSRQ), IMU (acc/gyr), GNSS quality indicators (SNR, satellite count) and system time tags including switch labels (e.g., WiFi$\rightarrow$Cell, AP handover). PDR provides a relative motion backbone; other modalities are time-aligned to this backbone and summarized into compact, privacy-preserving descriptors (IDs hashed; values normalized/quantized). The resulting temporal fingerprint at step/window $t$ is
\[
\mathbf{f}_t=\big[\phi_{\text{PDR}}\;\|\;\phi_{\text{WiFi}}\;\|\;\phi_{\text{Cell}}\;\|\;\phi_{\text{GNSS}}\;\|\;\phi_{\text{Time}}\;\|\;\mathbf{m}_t\big],
\]
where $\phi_{\cdot}$ are modality summaries and $\mathbf{m}_t$ encodes modality presence/quality. Switch labels are stored alongside $\mathbf{f}_t$ as weak event markers to anchor environment transitions and to seed prototype updates.

Fingerprints $\mathbf{f}_t$ accumulate over time to form a user-specific library on device, refreshed with simple aging/pruning to keep storage and compute low. This library supports real-time matching via AIDTW and exports only desensitized summaries to the cloud for policy updates in the cloud-edge RLHF loop.

\begin{figure}[t] %
  \centering
  \includegraphics[width=8.5cm]{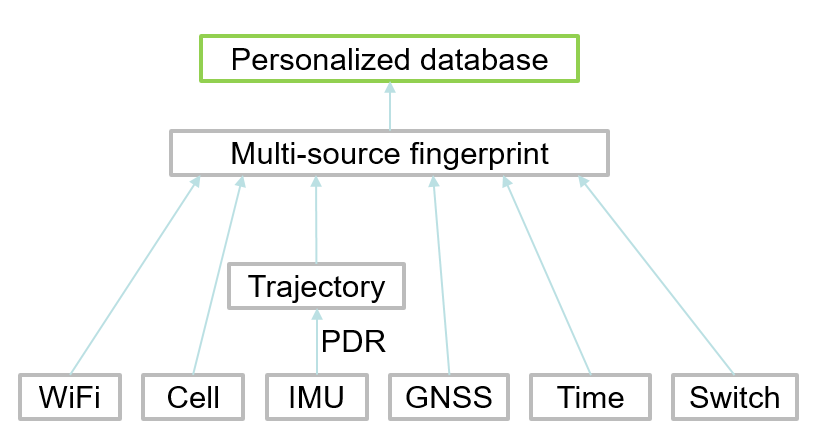}
  \caption{On-device construction of a personalized fingerprint library.}
  \label{fig:database}
\end{figure}

\subsection{AI-enhanced DTW (AIDTW)}
\label{sec:aidtw}

\begin{figure}[t]
  \centering
  \includegraphics[width=8.5cm]{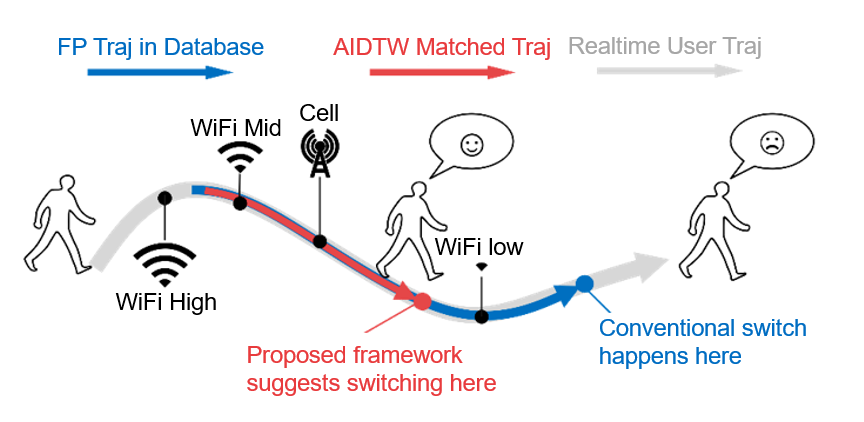} 
  \caption{AIDTW aligns the real-time trajectory (gray) to a fingerprint prototype (blue) and anticipates WiFi degradation; the policy suggests an earlier switch (red) than conventional methods (blue).}
  \label{fig:aidtw_switch}
\end{figure}

AIDTW augments DTW with two learnable components: (i) an adaptive filter selector that chooses which denoiser to use (Kalman / Gaussian / exponential low-pass) and its coefficients per window; (ii) a learned similarity used inside DTW with modality/time weights that emphasize WiFi and PDR.

\textbf{Learned adaptive filtering} Given window context $\mathbf{c}_t$ (RSSI variance, scan age, step rate, modality presence), a small network
$g_{\psi}(\mathbf{c}_t)$ outputs a distribution over filters and their parameters:
\[
\begin{aligned}
(\pi_t,\theta_t) &= g_{\psi}(\mathbf{c}_t), \pi_t \in \Delta^3,\\
\theta_t &\in \{\text{Kalman}(Q,R),\, \text{Gaussian}(\sigma),\, \text{ELP}(\alpha)\}.
\end{aligned}
\]
$g_{\psi}$ is trained to reduce Soft-DTW distance for true matches and increase it for non-matches; the network also regresses stable coefficients.

\textbf{Learned similarity inside DTW}, for cleaned query windows, AIDTW uses a modality-weighted, learned metric:
\[
d(i,j)=\sum_{m} \underbrace{w_m}_{\text{learned}}\,
\big\|\,W_m\,\tilde{\mathbf{x}}^{m,q}_i - W_m\,\tilde{\mathbf{x}}^{m,f}_j\big\|_2^2 \cdot \mathbf{1}_m(i,j),
\]
where $W_m$ (linear embedding) and $w_m$ (via softmax over trainable scores) are learned; $w_{\text{WiFi}},w_{\text{PDR}}\!\gg\!$ others in practice.

DTW with a Sakoe–Chiba band yields $D_{\text{AIDTW}}$ \cite{C7}; similarity is $S=\exp(-\beta D_{\text{AIDTW}})$ and $\beta$ is a learned inverse temperature for distance-to-similarity calibration. $\{W_m,w_m\}$ are trained jointly with $g_{\psi}$ using a margin loss with Soft-DTW so gradients flow through the alignment. 

\begin{figure}[htbp]
  \centering
  \includegraphics[width=8.5cm]{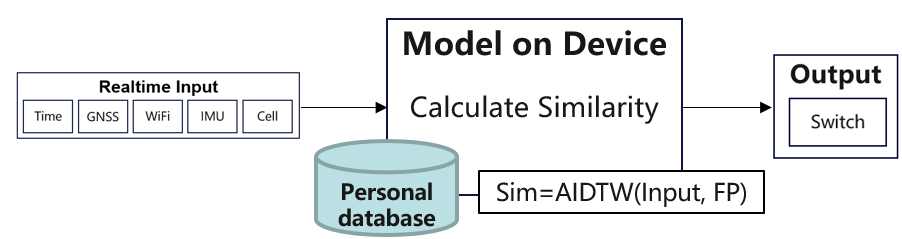} 
  \caption{AIDTW online similarity calculation process illustration.}
  \label{fig:aidtw_process}
\end{figure}

Weak supervision from switch tags form positive/negative pairs. The objective encourages filter choices and metrics that (i) denoise without delaying transitions and (ii) align WiFi/PDR trends that precede degradation. In deployment, the top-$S$ match anticipates low-WiFi segments along the user path (Fig.~\ref{fig:aidtw_switch}), enabling earlier policy actions.

\subsection{cloud-edge online RLHF for policy optimization}
To operationalize these capabilities, we replace rule-based decision logic with a cloud-edge collaboration framework for online RLHF. Edge devices perform on-device preprocessing and matching, generate privacy-preserving (desensitized) trajectory summaries, and execute a lightweight policy for environment-aware actions (e.g., scan pacing, pre-association, or connectivity selection) under latency/energy budgets. The cloud aggregates multi-user summaries to train (and periodically distill) a reward model and policy: a PPO agent maximizes a composite reward
\[
R \;=\; \eta\,\Delta\mathrm{Time} \;+\; \lambda\,\mathrm{Sim} \;+\; \gamma\,\mathrm{HF},
\]
where $\Delta\mathrm{Time}$ measures transition-time improvement, $\mathrm{Sim}$ measures calibrated matching quality from AIDTW, and $\mathrm{HF}$ encodes normalized human feedback (explicit confirmations/corrections and implicit signals such as rollbacks). The cloud-edge loop runs for $N$ rounds with periodic personalization and on-device policy updates (including offline RL when connectivity is limited), yielding fast adaptation without site-specific surveys.
\begin{figure}[htbp] 
  \centering
  \includegraphics[width=8.5cm]{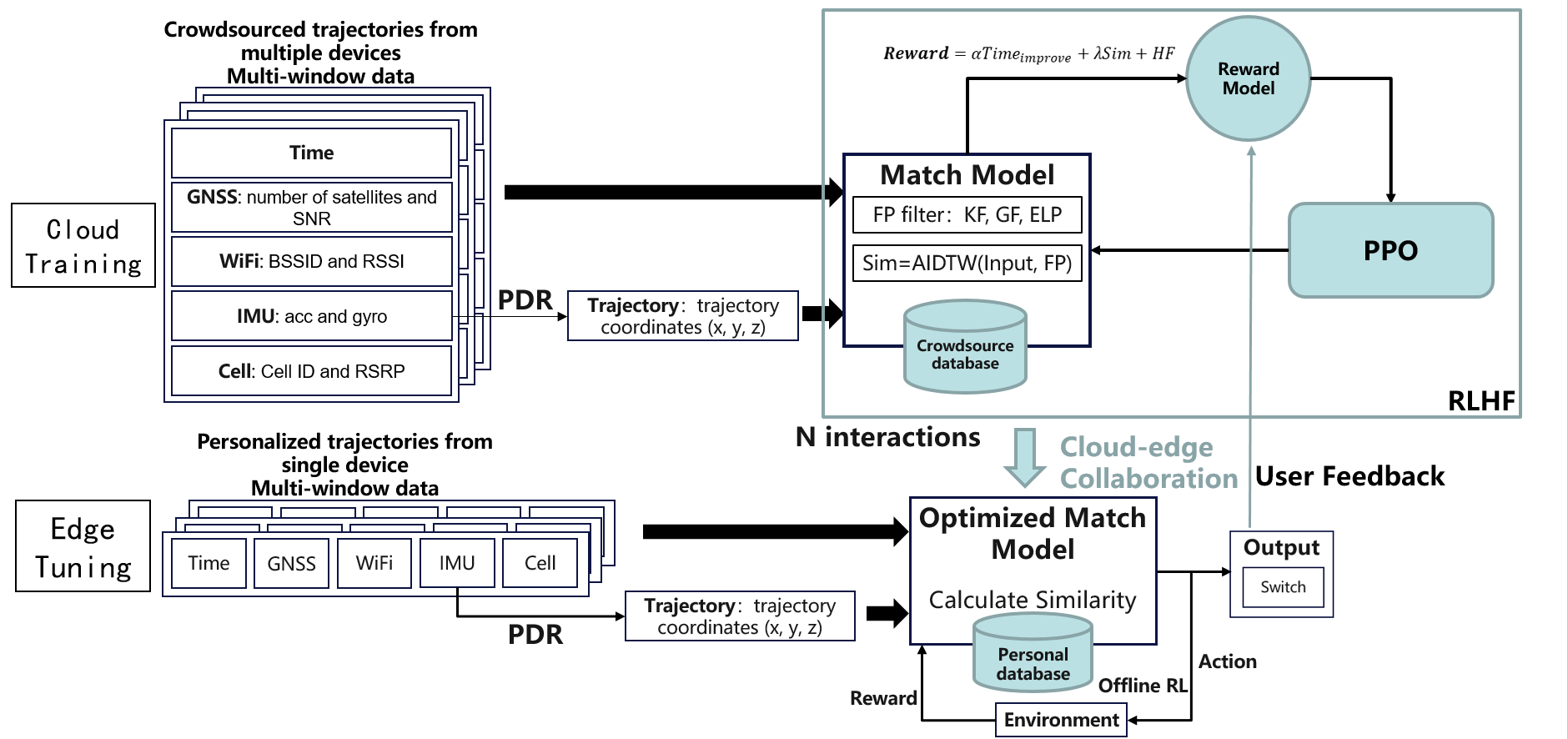}
  \caption{Cloud-edge RLHF: user feedback drives policy optimization and distillation.}
  \label{fig:cloudedge}
\end{figure}

\section{Experimental Setup}
\label{sec:exp-setup}

We evaluate in three real-world sites representative of common mobility patterns and short transitions that dominate perceived latency as shown below:Site A Office–Indoor: fully indoor paths within a business building; Site B Office–Door Egress: mostly indoor trajectories that exit through the front door of a business building and proceed briefly outside along the entrance apron; Site C Apartment–Mixed: paths starting inside an apartment building (corridor/stairwell), exiting to a courtyard/sidewalk, then walking outdoors for a short segment.
% \begin{figure}[!t] 
%   \centering
%   \includegraphics[width=8.5cm]{testing sites.png}
%   \caption{Testing sites illustrations: the arrows indicate moving directions.}
%   \label{fig:testsites}
% \end{figure}

Across sites, we collect multi-modal logs with a single commodity smartphone carried in hand or pocket at typical walking speeds. Trajectories include straight corridors, turns, short waits, and brief outdoor segments. All data are acquired during normal business/residential hours to capture realistic interference and crowding.

\section{Results and Analysis}
\label{sec:results}

\subsection{Site A—Office–Indoor}
\label{sec:siteA}

\begin{figure}[htbp]
  \centering
  \includegraphics[width=8.5cm]{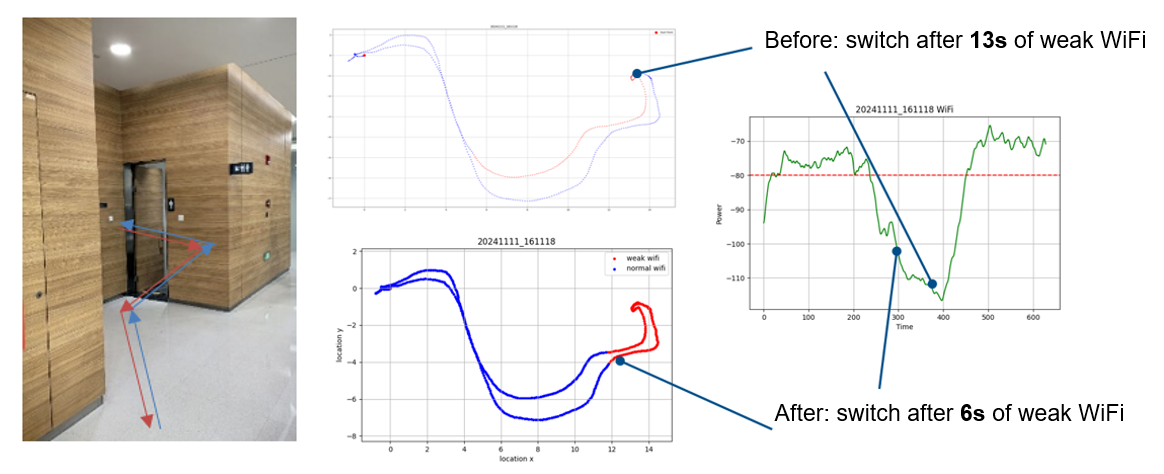}
  \caption{Site A restroom scenario. Left: restroom position within the office floor. Right: user trajectory with predicted low-WiFi segment and handover timing. Our method anticipates degradation and reduces TTS relative to a threshold baseline.}
  \label{fig:siteA_case}
\end{figure}

Figure~\ref{fig:siteA_case} illustrates a representative restroom-use scenario inside a business building. We highlight the difference between a heuristic baseline (RSSI threshold + hysteresis) and our method (AIDTW matching + RLHF policy): our policy initiates pre-association earlier and completes the handover closer to the true degradation onset. Below is the operational pipeline during the case (aligned with Fig.~\ref{fig:siteA_case}). 

\textbf{1. Multi-source fingerprinting (10\,s pre-switch buffer).} As the user approaches the restroom, PDR recognizes short stops/turns; WiFi scans show weakening BSSID; cellular/PCI remains stable; GNSS is mostly unavailable indoors. We buffer all modalities in the last 10\,s before a potential switch and commit the segment to the on-device fingerprint library only if a switch actually occurs (rolling retention 1–2 weeks). 

\textbf{2. Cloud-edge distillation with on-device personalization.} The edge runs a compact model distilled from a cloud pre-trained model and is further adapted by RLHF using the user’s personalized fingerprint library. 

\textbf{3. AI-enhanced online matching and decision.} AIDTW aligns the live window to the restroom-egress fingerprints (modality-weighted Soft-DTW with adaptive denoising) and produces a similarity score. When $S \ge \tau$, the policy triggers pre-association/scan pacing and, if needed, a proactive handover, yielding earlier, stable connectivity.

As shown in Table~\ref{tab:siteA_overall}, our method consistently lowers TTS with modest on-device overhead. The results show a consistent reduction in TTS across five sessions, with an average latency improvement of 6.95 s, confirming that our approach anticipates WiFi degradation earlier than baseline thresholds in fully indoor environments.

\begin{table}[htbp]
\centering
\small
\caption{Site A restroom scenario.}
\label{tab:siteA_overall}
\setlength{\tabcolsep}{6pt}
\begin{tabular}{lccccc}
\hline
 & \textbf{D1} & \textbf{Day2} & \textbf{Day3} & \textbf{Day4} & \textbf{Day5} \\
\hline
Baseline TTS (s) & 12.68 & 13.41 & 13.68 & 15.28 & 13.43 \\
Proposed TTS (s)     &  6.60 &  5.60 &  7.80 &  6.60 &  7.10 \\
Improvement (s)  &  6.08 &  7.81 &  5.88 &  8.68 &  6.33 \\
\hline
\textbf{Average (s)} & \multicolumn{5}{c}{\textbf{6.95}} \\
\hline
\end{tabular}
\vspace{0.25em}
\footnotesize
\emph{Notes.} The average relative improvement across these sessions is \(\approx 50.6\%\).
\end{table}

\subsection{Site B—Office Door Egress}
\label{sec:siteB}
\begin{figure}[htbp]
  \centering
  \includegraphics[width=8.5cm]{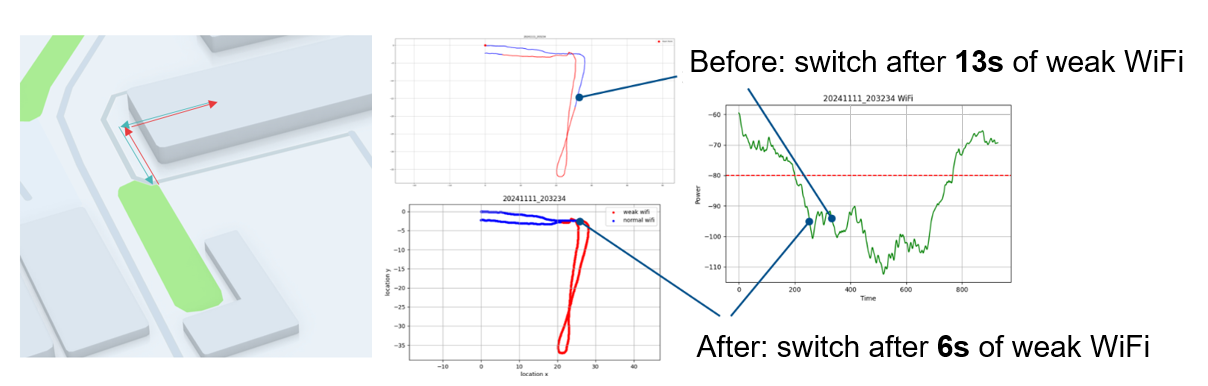}
  \caption{Site B door egress: the left panel shows the office entrance area; the right panel overlays the entry/exit trajectory and handover timing}
  \label{fig:siteB_case}
\end{figure}

Figure~\ref{fig:siteB_case} depicts the user leaving a business building. The process pipeline and presentation are identical to Site A section.

Although variability is higher in Table~\ref{tab:siteB_compact}, our method achieves an average 4.96 s absolute gain, demonstrating that proactive handover at door exits still significantly reduces latency despite occasional rebound effects near thresholds.

\begin{table}[htbp]
\centering
\scriptsize
\caption{Site B door egress scenario.}
\label{tab:siteB_compact}
\setlength{\tabcolsep}{2pt}
\renewcommand{\arraystretch}{1.12}
\begin{tabular}{lcccccccccc}
\hline
 & \textbf{D1} & \textbf{D2} & \textbf{D3} & \textbf{D4} & \textbf{D5} & \textbf{D6} & \textbf{D7} & \textbf{D8} & \textbf{D9} & \textbf{D10} \\
\hline
Baseline TTS (s) & 9.20 & 18.22 & 14.44 & 15.62 & 12.38 & 17.83 & 15.79 & 12.07 & 17.41 & 16.13 \\
Proposed TTS (s)     & 6.70 &  6.30 & 11.70 &  8.12 &  6.00 &  8.53 & 16.80 & 10.44 & 14.30 & 10.60 \\
Improvement (s)  & 2.50 & 11.92 &  2.74 &  7.50 &  6.38 &  9.30 & -1.01 &  1.63 &  3.11 &  5.53 \\
\hline
\textbf{Average (s)} & \multicolumn{10}{c}{\textbf{4.96}} \\
\hline
\end{tabular}
\vspace{0.25em}
\footnotesize
\emph{Notes.} The average relative improvement across these sessions is \(\approx 31.8\%\).
\end{table}

\subsection{Site C—Apartment–Mixed}
\label{sec:siteC}
Site C captures mixed indoor–outdoor transitions typical of apartment settings. A user begins inside the building (corridor, stairwell), exits through the main door, and continues walking outdoors in the courtyard and along the street. This scenario stresses the system’s ability to anticipate connectivity changes where GNSS signals reappear while WiFi rapidly weakens.

\begin{figure}[htbp]
  \centering
  \includegraphics[width=0.98\columnwidth]{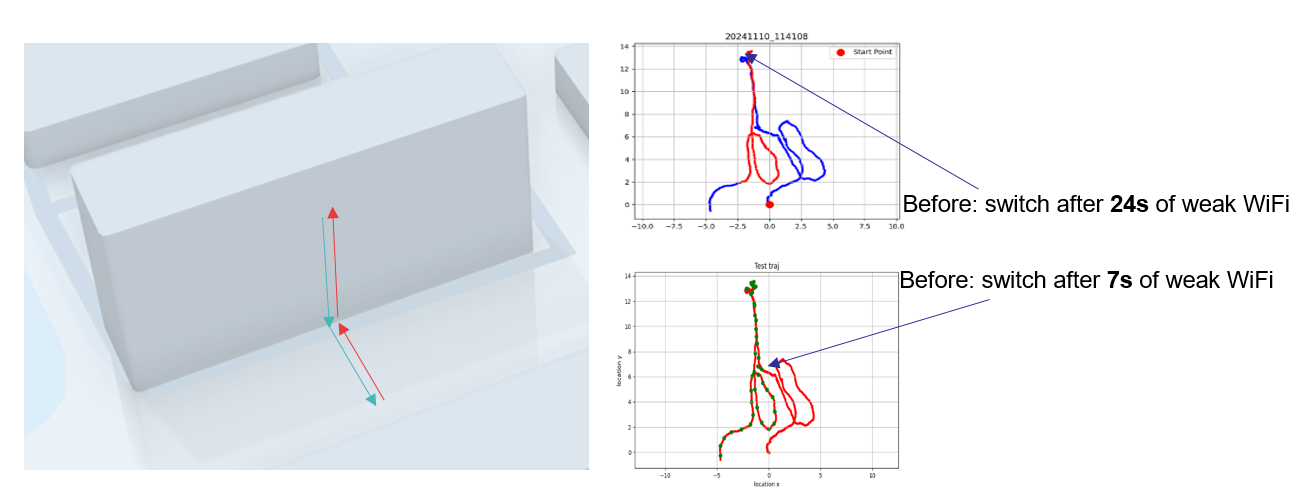}
  \caption{Site C mixed indoor–outdoor transition. Our method exploits GNSS reappearance + WiFi decay within 5\,s, enabling earlier handover than a threshold baseline.}
  \label{fig:siteC_case}
\end{figure}

Different from Site A and B, fingerprints are committed only when three conditions are jointly satisfied: (a) continuous high-confidence GNSS signals appear, (b) WiFi RSSI is weak or sharply decays within 5\,s, and (c) PDR indicates motion consistent with door exit. Figure~\ref{fig:siteC_case} shows one representative trajectory.

\begin{table}[htbp]
\centering
\scriptsize
\caption{Site C Apartment–Mixed scenario.}
\label{tab:siteC_outdoor}
\setlength{\tabcolsep}{3pt}
\renewcommand{\arraystretch}{1.12}
\begin{tabular}{lcccccc}
\hline
 & \textbf{Day1} & \textbf{Day2} & \textbf{Day3} & \textbf{Day4} & \textbf{Day5} & \textbf{Day6} \\
\hline
Baseline TTS (s) & 24.36 & 20.33 & 19.31 & 17.54 & 15.56 & 20.51 \\
Proposed TTS (s)     &  7.70 &  6.30 &  6.50 &  7.70 &  5.30 &  7.10 \\
Improvements (s) & 16.66 & 14.03 & 12.81 &  9.84 & 10.26 & 13.41 \\
\hline
\textbf{Average (s)}    & \multicolumn{6}{c}{12.80} \\
\hline
\end{tabular}
\vspace{0.25em}
\parbox{\columnwidth}{\footnotesize\emph{Notes.} The average relative improvement across these sessions is \(\approx 65.2\%\).}

\end{table}

In Table~\ref{tab:siteC_outdoor}, gains are most pronounced: an average 12.8 s reduction, reflecting the strong complementary value of GNSS reappearance and WiFi decay signals during mixed indoor–outdoor transitions.

\section{Conclusion and Future Work}
We present a novel AI-enhanced framework for personalized, survey-free environment recognition using sensor fusion, without site-specific pre-deployment. The framework is modular: IMU–PDR trajectories can be augmented by Simultaneous Localization and Mapping (SLAM), while the essential principle is combining spatial trajectories with signal patterns to capture environmental effects. Propagation-aware simulation can synthesize environment-conditioned fingerprints for training and calibration \cite{C22}. Future work will pursue semantic scene labeling, device adaptation, and federated personalization to ensure scalability and privacy.

\clearpage

\bibliographystyle{IEEEbib}
\bibliography{strings,refs}

\begin{thebibliography}{10}

\bibitem{C1}
P.~Bahl and V.N. Padmanabhan,
\newblock ``Radar: An in-building rf-based user location and tracking system,''
\newblock in {\em Proceedings of IEEE INFOCOM}. IEEE, 2000, vol.~2, pp.
  775--784.

\bibitem{C2}
M.~Youssef and A.~Agrawala,
\newblock ``The horus wlan location determination system,''
\newblock in {\em Proceedings of the 3rd International Conference on Mobile
  Systems, Applications, and Services (MobiSys)}. USENIX, 2005, pp. 205--218.

\bibitem{C3}
B.~Ferris, D.~H{\"a}hnel, and D.~Fox,
\newblock ``Gaussian processes for signal strength-based location estimation,''
\newblock in {\em Proceedings of Robotics: Science and Systems (RSS)}. MIT
  Press, 2006.

\bibitem{C16}
K.~Wu, J.~Xiao, Y.~Yi, D.~Chen, and L.M. Ni,
\newblock ``Csi-based indoor localization,''
\newblock in {\em Proceedings of IEEE INFOCOM}. IEEE, 2012.

\bibitem{C4}
X.~Wang, L.~Gao, S.~Mao, and S.~Pandey,
\newblock ``Deepfi: Deep learning for indoor fingerprinting using channel state
  information,''
\newblock in {\em Proceedings of IEEE WCNC}. IEEE, 2015.

\bibitem{C14}
X.~Wang, X.~Wang, S.~Mao, J.~Zhang, S.C.G. Periaswamy, and J.~Patton,
\newblock ``Indoor radio map construction and localization with deep gaussian
  processes,''
\newblock in {\em IEEE Internet of Things Journal}. IEEE, 2020.

\bibitem{C5}
C.~Wu, Z.~Yang, C.~Xiao, C.~Yang, Y.~Liu, and M.~Liu,
\newblock ``Self-updating radio maps for wireless indoor localization,''
\newblock in {\em Proceedings of IEEE INFOCOM}. IEEE, 2015.

\bibitem{C13}
W.~Li, D.~Wei, Q.~Lai, X.~Li, and H.~Yuan,
\newblock ``Geomagnetism-aided indoor wi-fi radio-map construction via
  smartphone crowdsourcing,''
\newblock in {\em Sensors}. MDPI, 2018.

\bibitem{C15}
Y.~Sun, Y.~Yang, J.~Li, Y.~Liu, and Z.~Wang,
\newblock ``Voronoi diagram and crowdsourcing-based radio map interpolation for
  fingerprinting localization,''
\newblock in {\em IEEE Access}. IEEE, 2018.

\bibitem{C18}
Anshul Rai, Krishna~Kant Chintalapudi, Venkata~N. Padmanabhan, and Rijurekha
  Sen,
\newblock ``Zee: Zero-effort crowdsourcing for indoor localization,''
\newblock in {\em Proceedings of the 18th Annual International Conference on
  Mobile Computing and Networking (MobiCom)}. ACM, 2012, pp. 293--304.

\bibitem{RIO2022}
Xiya Cao, Caifa Zhou, Dandan Zeng, and Yongliang Wang,
\newblock ``Rio: Rotation-equivariance supervised learning of robust inertial
  odometry,''
\newblock in {\em Proceedings of the IEEE/CVF Conference on Computer Vision and
  Pattern Recognition (CVPR)}, 2022, p. 650–660,
\newblock Also available on arXiv:2111.11676.

\bibitem{C12}
J.~Haverinen and A.~Kemppainen,
\newblock ``Global indoor self-localization based on the ambient magnetic
  field,''
\newblock in {\em Robotics and Autonomous Systems}. Elsevier, 2009.

\bibitem{C7}
E.~Keogh and C.A. Ratanamahatana,
\newblock ``Exact indexing of dynamic time warping,''
\newblock in {\em Proceedings of the 31st International Conference on Very
  Large Data Bases (VLDB)}. VLDB Endowment, 2005, pp. 406--417.

\bibitem{C6}
M.~Cuturi and M.~Blondel,
\newblock ``Soft-dtw: A differentiable loss function for time-series,''
\newblock in {\em Proceedings of the 34th International Conference on Machine
  Learning (ICML)}. PMLR, 2017, pp. 894--903.

\bibitem{C19}
Chien-Yi Chang, De-An Huang, Yanan Sui, Fei-Fei Li, and Juan~Carlos Niebles,
\newblock ``D3tw: Discriminative differentiable dynamic time warping for weakly
  supervised action alignment and segmentation,''
\newblock in {\em Proceedings of the IEEE/CVF Conference on Computer Vision and
  Pattern Recognition (CVPR)}. IEEE/CVF, 2019, pp. 3546--3555.

\bibitem{C11}
J.~Schulman, F.~Wolski, P.~Dhariwal, A.~Radford, and O.~Klimov,
\newblock ``Proximal policy optimization algorithms,''
\newblock in {\em arXiv preprint arXiv:1707.06347}. arXiv, 2017.

\bibitem{C10}
P.F. Christiano, J.~Leike, T.B. Brown, M.~Martic, S.~Legg, and D.~Amodei,
\newblock ``Deep reinforcement learning from human preferences,''
\newblock in {\em Proceedings of the 31st Conference on Neural Information
  Processing Systems (NeurIPS)}. NeurIPS, 2017.

\bibitem{C8}
H.B. McMahan, E.~Moore, D.~Ramage, S.~Hampson, and B.A. y~Arcas,
\newblock ``Communication-efficient learning of deep networks from
  decentralized data,''
\newblock in {\em Proceedings of the 20th International Conference on
  Artificial Intelligence and Statistics (AISTATS)}. PMLR, 2017, pp.
  1273--1282.

\bibitem{C9}
G.~Hinton, O.~Vinyals, and J.~Dean,
\newblock ``Distilling the knowledge in a neural network,''
\newblock in {\em Proceedings of the NIPS Deep Learning and Representation
  Learning Workshop}. NeurIPS, 2015.

\bibitem{C20}
Timo Kaufmann, Paul Weng, Viktor Bengs, and Eyke H{\"u}llermeier,
\newblock ``A survey of reinforcement learning from human feedback,''
\newblock {\em arXiv preprint arXiv:2312.14925}, 2023.

\bibitem{C22}
Ruichen Wang, Samuel Audia, and Dinesh Manocha,
\newblock ``Indoor wireless signal modeling with smooth surface diffraction
  effects,''
\newblock in {\em Proceedings of the 18th European Conference on Antennas and
  Propagation (EuCAP)}. IEEE, 2024, pp. 1--5.

\end{thebibliography}

\end{document}